\documentclass[acmsmall,screen,nonacm,review=false,timestamp=false]{acmart}
\usepackage[utf8]{inputenc}
\AtBeginDocument{%
  }

\usepackage{fancyhdr}
\AtBeginDocument{%
    \addtolength{\footskip}{1.0\baselineskip}%
    \fancyfoot[L]{\textit{\textbf{Preprint --- please check authors' websites for updated versions.}}}%
}

\definecolor{maria-color}{HTML}{7881F2}

\setcopyright{acmlicensed}
\copyrightyear{2026}
\acmYear{2026}
\acmDOI{XXXXXXX.XXXXXXX}

\acmConference[FAccT '26]{Ninth Annual ACM Conference on Fairness, Accountability, and Transparency}{Montreal, CA}

\begin{document}

\title{Provocations from the Humanities for Generative AI Research}


\author{Lauren Klein}
\affiliation{%
  \institution{Emory University}
  \city{Atlanta}
  \state{GA}
  \country{USA}
}
\email{lauren.klein@emory.edu}

\author{Meredith Martin}
\affiliation{%
  \institution{Princeton University}
  \city{Princeton}
  \state{NJ}
  \country{USA}
}
\email{mm4@princeton.edu}

\author{Andre Brock}
\affiliation{%
  \institution{Georgia Institute of Technology}
  \city{Atlanta}
  \state{GA}
  \country{USA}
}
\email{andre.brock@lmc.gatech.edu}

\author{Maria Antoniak}
\affiliation{%
  \institution{University of Colorado Boulder}
  \city{Boulder}
  \state{Colorado}
  \country{USA}
}
\email{maria.antoniak@colorado.edu}

\author{Melanie Walsh}
\affiliation{%
  \institution{University of Washington}
  \city{Seattle}
  \state{WA}
  \country{USA}
}
\email{melwalsh@uw.edu}

\author{Jessica Marie Johnson}
\affiliation{%
  \institution{Johns Hopkins University}
  \city{Baltimore}
  \state{MD}
  \country{USA}
}
\email{jmj@jhu.edu}

\author{Lauren Tilton}
\affiliation{%
  \institution{University of Richmond}
  \city{Richmond}
  \state{VA}
  \country{USA}
}
\email{ltilton@richmond.edu}

\author{David Mimno}
\affiliation{%
  \institution{Cornell University}
  \city{Ithaca}
  \state{NY}
  \country{USA}
}
\email{mimno@cornell.edu}

\renewcommand{\shortauthors}{Klein et al.}

\begin{abstract}
The effects of generative AI are experienced by a broad range of constituencies, but the disciplinary inputs to its development have been surprisingly narrow. Here we present a set of provocations from humanities researchers---currently underrepresented in AI development---intended to inform its future applications and enrich ongoing conversations about its uses, impact, and harms. Drawing from relevant humanities scholarship, along with foundational work in critical data studies, we elaborate eight claims with broad applicability to generative AI research: 1) Models make words, but people make meaning; 2) Generative AI requires an expanded definition of culture; 3) Generative AI can never be representative; 4) Bigger models are not always better models; 5) Not all training data is equivalent; 6) Openness is not an easy fix; 7) Limited access to compute enables corporate capture; and 8) AI universalism creates narrow human subjects. We also provide a working definition of humanities research, summarize some of its most salient theories and methods, and apply these theories and methods to the current landscape of AI. We conclude with a discussion of the importance of resisting the extraction of humanities research by computer science and related fields.  	
\end{abstract}



\received[submitted]{13 January 2025}
\received[revised]{12 January 2026}

\maketitle

\section{Introduction}
Nine years after the introduction of the transformer, four years after the first public release of ChatGPT, and one year after Elon Musk brought (and then abandoned) DOGE to the US federal government, we are at an inflection point with respect to generative AI. Its harms to people and to the planet are well documented. Its threats to media ecosystems---and to any number of creative industries---are self-evident. And its dangers to society and governance, those brought about by unchecked corporate decision-making, are crystal clear. We write as humanities researchers with long histories of engaging with computational technologies, particularly through the field of digital humanities (DH), who share these concerns, and seek to contribute our expertise to the alternative ecosystem of generative AI research that is actively in formation at FAccT and allied venues. More concretely, we seek to redress what philosopher Charles Mills might describe as the ``epistemology of ignorance'' that dominates corporate AI research \cite{mills_racial_1999}. This is the worldview that results from an intentional avoidance of frameworks that challenge normative beliefs, and serves only to reinforce existing dominant perspectives. We see this especially in papers making broad claims about language, culture, and values, as well as in systems designed to serve a universal user or ``fix'' a complex social issue. Even work that purports to engage with humanistic ways of thinking tends to demonstrate a limited engagement with work coming out of actual humanities disciplines---English and other literary fields, history and art history, American, African American, and ethnic studies, musicology and philosophy, women’s and gender studies, film and media studies, and more. As a result, the deep expertise that humanities researchers possess---our knowledge about the past, our ability to conduct detailed analyses within and between cultures, and our command of meaning-making practices past and present, among others---has not been leveraged in generative AI research.  

We contend that this omission is, in large part, the result of disciplinary differences in research methods and outputs. In the humanities, we build up systems-level analyses from specific objects of culture. A poem or an artwork, a community on Twitter, an interview with a local elder, a historical newspaper or court transcript, or even a large language model, might function as the focal point of an analysis that leads to an interpretation, and in turn to what we call an interpretive model. In the humanities, we call these models ``theories,'' but not in the sense that technical disciplines use this term. Our theories, grounded in specificity, help us better understand, navigate, and posit additional answers to questions that are otherwise too big, too complex, or too intractable to be able to resolve with any precision. The advent of generative AI requires exactly this kind of work.  

In this paper, we offer a set of provocations for generative AI research that are informed by these theories, and other areas of humanistic expertise. In structuring our contribution, we draw from the framework developed by danah boyd and Kate Crawford (\citeyear{boyd_critical_2012}) in ``Critical Questions for Big Data,'' in which they present a similar set of provocations for what was then the new landscape of Big Data. Inspired by the transformative impact of their work, we offer eight new provocations for the landscape of generative AI. They are as follows: 1) Models make words, but people make meaning; 2) Generative AI requires an expanded definition of culture; 3) Generative AI can never be representative; 4) Bigger models are not always better models; 5) Not all training data is equivalent; 6) Openness is not an easy fix; 7) Limited access to compute enables corporate capture; and 8) AI universalism creates narrow human subjects. Each provocation consists of an assertion that is elaborated through current humanities research, and is intended to encourage AI researchers---including those at FAccT---to more fully consider how their work could be enriched by incorporating humanistic expertise. Each provocation concludes with recommendations for next steps.

\section{Background: What are the humanities anyway?}
The humanities have an extensive institutional and disciplinary history \cite{bod_new_2016, noauthor_philology_nodate, reitter_permanent_2021}; here we focus on humanities research as it is practiced in the present. To conduct humanities research entails an investigation of the human---of people and groups, 
and
the cultural objects they create \cite{small_introduction_2013,haufe_humanities_2024}. The primary goal of this work is to understand how these cultural objects create or reflect new forms of meaning and knowledge. To do so, we are trained in the history of knowledge production. We are also trained to translate cultural and historical meaning across objects and audiences, often in multiple languages, as well as in and across a variety of forms, including digital forms. 

Humanities research requires minute specificity and sweeping breadth. For example, to analyze Jamaican-American poet Claude McKay's 1912 book \textit{Constab Ballads}  \cite{singh_mckay_scalar}, a scholar must first learn to read both normative English and McKay's version of a historical Jamaican dialect filled with elisions and diacritical marks. They must be able to analyze McKay's poetry in the context of Caribbean history, American literature, and Black studies. They must understand both the material culture that informs the book's original printed form and the choices that inform the text encoding and platform development when they encounter the work online, as most scholars do today. They must navigate the work's relationship to dialect poetry in the early 20th century, as well as to questions of canon formation, past and present. In this case and more broadly, humanities researchers are trained to understand concepts by examining and interpreting a specific case, and by asking what ways of knowing and thinking (individually and collectively) might have gone into its creation \cite{singh_mckay_scalar}. Thus, in addition to an array of interpretive strategies, humanities research entails an attention to broader concepts such as language, history, philosophy, theory, and aesthetics. 

What methods are employed to conduct this work? Methods shared across the humanities  include interpretation, contextualization, and theorization, as just described. Specific disciplinary techniques include textual explication and close reading (literary studies), historical synthesis (historical fields), and media-specific analysis (film and media studies), among many others. Humanities scholars may also visit archives, interview artists, or compile data or metadata. These methods---some that take place in the mind, and others in the world---are what generate the bulk of the evidence that appears in humanities scholarship. However, what more commonly travels to technical fields are the theories that this evidence points towards. These are more precisely described as humanistic theories: written articulations of complex ideas that, until that point, we did not yet fully understand (or we thought we did, but evidence shows there is still more to learn). In this paper, we summarize and synthesize many of these theories. Readers who want to learn more about the methods and mechanisms of humanities research should consult our bibliography. 

The end result of humanities research is very rarely a simple conclusion. Rather, the goal is a deeper understanding of the ``big question'' that frames the work. 
This is what classicist Gregory Crane, in the context of a discussion of the relationship between AI and the humanities, describes as an augmentation of \textit{our own} intelligence \cite{Crane_2019} (emphasis added). What humanities researchers can contribute to the current conversation about generative AI, its present uses, its incontrovertible harms, and its future possibilities, then, is additional clarity and expertise-driven specificity about the stakes of engaging with AI development in an age of techno-capital, and what it means for the human record. This contribution is generative in the original sense of the word: it enables our work with (or against) AI to progress because it allows us to conduct more informed, more accurate, and more humane (in the sense of ``people-centered'') research. 

As should be clear, we reject formulations of humanistic thinking that have been proffered by scholars in computer science as, for example, ``basically a style or mode'' \cite{Bardzell_Bardzell_2015}. Our methods are instilled through rigorous training, refined through ongoing, reflection-driven interpretation, and grounded in the history of how an object or culture or concept has been interpreted to that point. Like the sciences, the humanities are iterative; humanities scholars build on past knowledge, and humanities disciplines grow and they change. The impact of this work is substantial. The humanities teach us what specific cultural objects (not limited to aesthetic objects but all human production, including science and technology) might have meant to the people who created them or used them, and about the cultures that gave rise to them. These objects and cultures, in turn, point to how humans have existed in cultural, social, and sociotechnical contexts past and present, and how these contexts are part of what defines us at any given moment. 

Key to this work is an understanding of how these stories may consist of that which has not yet been discovered, digitized, transcribed, or annotated, as well as the of unrecoverable voices, the silenced narratives, and the stolen artifacts that have become symbols of empire. As part of this effort, like the social sciences, we also study technologies to understand how they affect societies and cultures \cite{Hicks_2020, Morgan_2021, Rosenthal_2018, benjamin_imagination_2024} and how they allow for new forms of cultural expression and representation, and we use technologies to tell new stories and make new discoveries about the past \cite{gallon_chapter_2016}. We have learned how interpretations of the past become the subject of future interpretations, which is why we are so invested in informing the conceptual underpinnings of generative AI, its inputs, and its outputs, with humanistic expertise. In this spirit, we position our provocations as bridges and not walls. It is our collective belief that, just as we need a shared and informed understanding of what is possible (and also, importantly, not possible) with generative AI, we also need a more informed understanding of the cultures that enabled its development, the groups that might benefit from its future applications, and those that currently bear its costs.

\section{Related Work}

\subsection{Humanities Research at FAccT}
Historically, humanities research has not featured prominently within the ACM Conference on Fairness, Accountability, and Transparency (FAccT). Most mentions of the humanities in FAccT publications appear under the umbrella term of ``social science and humanities research,'' and the ACM category of ``arts and humanities'' also includes papers on ML, AI, and artistic practice, making core humanistic contributions difficult to surface. 
The main exception to this claim is the field of philosophy, which is one of the core disciplines of the FAccT community. A recent retrospective of the FAccT conference categorized 11\% of papers presented at FAccT as belonging to philosophy, and philosophy was the only humanities discipline prominent enough to be categorized individually \cite{Laufer_Jain_Cooper_Kleinberg_Heidari_2022}. 

Why is philosophy privileged at FAccT, as it is in public discourse about AI? We contend that, unlike other humanities disciplines, which prioritize the specific and the contextual, the field of (analytic) philosophy encourages researchers to generalize human responses and behaviors, or, alternately, isolate specific variables of interest in the service of a broader claim. This approach has resulted in significant contributions---both to FAccT and to the field of AI. Our line of questioning is intended to underscore how the humanities have more to offer the FAccT community than the contributions of one branch of philosophy alone. Indeed, it might be noted how philosophical approaches often (but not always e.g. \cite{Fazelpour2020,Sober2022-SOBPIS,TolbertForthcoming-TOLAAA-6}) conveniently mirror the engineering impulse to propose a technical ``fix'' to a complex social problem. As such, the uplifting of philosophy---and not the full range of humanistic contributions---may offer an all too tempting proposition for technical researchers who value more tractable theories of knowing. We can see this pattern reflected in works published at FAccT that have brought philosophical approaches to algorithmic decision-making \citep{Cooper_Moss_Laufer_Nissenbaum_2022, Susser_2022, Jain_Suriyakumar_Creel_Wilson_2024}. 

Nevertheless, concepts and ideas from the humanities (if not substantive humanities research) have consistently made their way to FAccT, often in impactful ways. In 2020, \citet{Pritchard_Snodgrass_Morrison_Britton_Moll_2020} organized a CRAFT workshop on speculative, queer, and feminist engagements with archives, and \citet{Jo_Gebru_2020} published ``Lessons from Archives: Strategies for Collecting Sociocultural Data in Machine Learning,'' a paper that has only grown in significance as the issue of LLM training data has entered broad consciousness. (\citet{croskey-2025}'s 2025 paper on liberatory collections extends this work in important ways.) Additional papers have focused on specific theoretical constructs from the humanities, using them to question core concepts in ML/AI research (e.g. Corbett and Denton on transparency \cite{Corbett_Denton_2023} and Stark on animation \cite{Stark_2024}), and to imagine alternatives to existing approaches (e.g. \citet{Klumbyte_Draude_Taylor_2022}, who explore the feminist and Black feminist concepts of situatedness, figuration, diffraction, and critical fabulation to think beyond existing modeling approaches). Feminist and gender theories, which extend into the social sciences, have received increasing attention at FAccT \cite{Devinney_Björklund_Björklund_2022, Klein_D’Ignazio_2024, valdivia_10.1145/3715275.3732027}. 
Engagement with art history and art practice can be found as well \cite{Huang_Liem_2022,Divakaran_Sridhar_Srinivasan_2023, Srinivasan_2024}.

There has also been sustained recognition of the distinctive perspectives that humanities (and social science) researchers bring to AI research and education, reflected in interdisciplinary workshops on AI fairness and explainability and in pedagogical work on integrating critical and responsible AI approaches into the classroom \cite{Ganesh_Dechesne_Waseem_2020, Lünich_Keller_2024, Bates_2020, Dotan_Parker_Radzilowicz_2024}.
Most pointedly, and ironically, Raji et al.’s quantitative analysis of AI ethics syllabi across the academy, ```You Can’t Sit With Us’: Exclusionary Pedagogy in AI Ethics Education'' \cite{Raji_Scheuerman_Amironesei_2021}, makes explicit mention of multiple humanities fields in the service of an argument about how, in refusing to look outside the discipline of computer science for their assigned readings, AI ethics courses are reproducing existing academic hierarchies.  We build upon this work by documenting the core methods of humanities research and synthesizing the major concepts that carry across its fields, showing how these methods and concepts might enrich future work at FAccT and generative AI research more broadly. 

\subsection{Humanities Research on AI}
The launch of ChatGPT, in November 2022, brought humanities researchers to the forefront of national conversations about generative AI and its impact on research, writing, and creativity \cite{schmidt_representation_2023,tilton_relating_2023,jones_ai_2023,crawford_archeologies_2023,broussard2023,underwood_mapping_2022}. In particular, those in the fields of digital humanities, digital pedagogy, and the history of technology---fields requiring both technical and humanistic expertise---became some of the earliest expositors of its limits and potential \cite{Kirschenbaum_2023, Rogers_2023, Klein_2022}. While these fields continue to lead within the humanities---for example, the 2023 Modern Language Association (MLA) and Conference on College Composition and Communication (CCCC) task force, led by digital humanists, published a set of working papers on AI and writing \cite{antoniobyrdMLACCCCJointTask}---the national conversation about AI has been usurped by corporate spokespeople, whose claims in turn convince academic administrators to adopt costly and potentially harmful AI systems \cite{cottom_2025}. In fact, the increasingly politicized relationship between for-profit corporations, public resources, public institutions (like universities and research centers), and political power should alarm humanists and computer scientists alike.  

Humanities researchers have also established new journals like \textit{Critical AI} \cite{Goodlad_2023} and published themed sections and special issues of flagship humanities journals 
\cite{Raley_Rhee_2023} like \textit{PMLA} 
\cite{Kirschenbaum_Raley_2024}. Monographs \citep{Pasquinelli, Gunkel_2012, Katz_2020, Tenen_2024} and long-form essays theorize the language \cite{Slater_2024} and images \cite{Wasielewski_2024} generated by LLMs, historicize and critique ML approaches such as sentiment analysis \cite{Chun_Hong_Nakamura_2024}, topic modeling \cite{Binder_2016}, and image generation \cite{Offert_Phan_2024}, analyze the output of generative AI according to literary critical \cite{Walsh_Preus_Gronski_2024} and art historical expertise \cite{Malevé_Sluis_2023}. Yet this valuable work is rarely read or cited beyond humanities fields. The goal of this paper is to argue that this work should have a direct impact on the development of generative AI. 

\section{Provocations from the Humanities for Generative AI Research}

We employ \citet{boyd_critical_2012}'s framework of ``provocations,'' developed with respect to Big Data, to formulate eight new provocations for generative AI research. These are intended to establish a stronger foundation for generative AI research and cross-disciplinary collaboration, to allow for the development of more informed and intentional research questions, and to prevent non-experts from making thin claims about the full breadth of human culture. We list the full set of provocations and suggested next steps in Table \ref{table:provocations-list}.

\begin{table}[h]
\centering
\footnotesize
\renewcommand{\arraystretch}{1.5}  
\setlength{\tabcolsep}{8pt}        
\begin{tabular}{|p{1cm}|p{5.3cm}|p{5.8cm}|}
\hline
\textbf{Keyword} & \textbf{Provocation} & \textbf{Suggested Next Steps} \\
\hline
Language & Models make words, but people make meaning & Familiarize yourself with theories of language and meaning beyond the linguistic frameworks that inform NLP. \\
\hline
Culture & AI requires an understanding of culture & Clarify which definition of culture you are using: ``culture'' as a group of people, or as the objects or expressions of a group of people. \\
\hline
Archive & AI can never be ``representative'' & Abandon the chase for a ``representative'' dataset and instead, acknowledge that there are always perspectives encoded in datasets and models. \\
\hline
Model & Bigger models are not always better models & Begin by asking whether a small model might be better for the task, and if so, involve domain experts. \\
\hline
Data & Not all training data is equivalent & Consider pretraining data's conceptual and cultural characteristics in addition to its technical characteristics and ``impact'' on model performance. \\
\hline
Access & Openness is not an easy fix & Ask whose knowledge your data or model is attempting to represent, whether you are authorized to make use of it, and whether your work might result in future harm. \\
\hline
Capture & Limited access to compute enables corporate capture & Identify your role in the larger mechanism of capitalism and use that knowledge to ask how to build futures that benefit more than the rich few. \\ 
\hline
Human & AI universalism creates narrow human subjects & Acknowledge the long history of abstraction and extraction so as to imagine more liberatory futures. \\
\hline
\end{tabular}
\caption{The full list of provocations and suggested next steps.}
\label{table:provocations-list}
\end{table}

\subsection{Models make words, but people make meaning}

What is the significance of language that is produced in a way that is---by construction---devoid of human intention? The field of natural language processing (NLP) has been shaped by decades of research in both machine learning and linguistics. This has led to explanations (and rejections) of the language generated by LLMs and related AI systems that rely on theories from the field of linguistics about ``communicative intent'' \cite{Bender_Gebru_McMillan-Major_Shmitchell_2021}. To be clear: framing model output as intentionless is essential for understanding its significance (or the lack thereof) in conversational contexts. But we can use additional theories from the humanities---and from other strains of linguistics---to understand the broader significance (or the lack thereof) of the language these models produce. 

More concretely, the theories of meaning-making (in both semantics and semiotics) developed by linguistic and literary scholars in the mid-twentieth century show how all sign-systems, including but not limited to language, have non-referential (i.e. non-mimetic) qualities that are nevertheless meaningful, and may serve functions beyond the merely communicative  \cite{jakobson1987language, barthes1977image, Derrida_1998, Cixous1976, spillers_mamas_1987}. This work, which is increasingly being taken up by scholars working at the intersection of literary theory and AI \cite{geogheganCodeInformationTheory2022, Tenen_2024, weatherby-2025}, offers some ways of deriving meaning from language that are not dependent upon being able to identify any particular speaker or source. For example, philosopher Roland Barthes proposed that the meaning of any particular text is determined not by the author’s intention, but by the reader’s interpretation \cite{barthes1977image}. Scholars in the fields of Black studies, ethnic studies, gender and sexuality studies, have critiqued the literal grammar of the English language, as that which  has long functioned to deny people their humanity \cite{spillers_mamas_1987}. Far from enabling a descent into relativism, theories such as these are the basis for an expanded concept of meaning, of the communities that create meaning and for what reasons, and of techniques we can employ (or ourselves develop) for making sense of the range of writing, and objects, we encounter in the world. 

LLMs generate text by predicting sequences of words based on both observed patterns and on human preferences and feedback. The result may be output that is factually wrong yet linguistically fluent and seemingly coherent---the ``hallucinations'' that have become a topic of research \cite{koenecke_careless_2024} and justifiable concern. The theories above allow us to understand how the human tendency to create meaning through the interpretive expectations of intention and care can confer the illusion that the model ``knows'' something or someone. The current climate of ``AI hype'' and the nonconsensual infusion of AI into our lives have naturalized artificial ``knowing'' over the complicated, messy, and polyrhythmic meaning-making conducted by humans when using language (oral, written, and otherwise) in social contexts. Humanists have been at the forefront of analyzing the output of generative AI models not for what they ``know,'' but for what meaning their output elicits about the cultures from which they emerged \cite{sui_confabulation_2024, Gunkel2025différance, handman-2026}. 

Meaning is always people-fueled, socially-driven, and impossible to map consistently. Especially in the context of the Global Majority, making meaning from language requires an awareness of the history of evading and erasing dominant meanings, and extracting additional meaning from guarded words. The idea of AI ``knowing'' is a false solution to the more complicated reality of meaning that is forged between language, place, time, and social relations. \textbf{Our suggestion for AI researchers is to familiarize themselves with theories of language and meaning beyond the linguistic frameworks that inform NLP, as well as histories of language being used as mechanisms for controlling populations and as weapons of imperial power.} These histories and theories can point to additional places in the text (or image or sound or multimedia) generation pipeline where meaning is made and where meaning is falsely attributed, as well as to the people for and about whom this meaning matters. 


\subsection{AI requires an understanding of culture}

Several years after the release of generative AI models to the public, we have ample evidence of how the text and images that they produce do not perform well in the context of non-dominant cultures, and at times actively harm them  \cite{Bender_Gebru_McMillan-Major_Shmitchell_2021,Prabhakaran_Qadri_Hutchinson_2022,cetinic_myth_2022,Cao_Zhou_Lee_Cabello_Chen_Hershcovich_2023}. Admirably, the technical community has recognized the problem and has proposed solutions such as improved training data \cite{Pawar_Park_Jin_Arora_Myung_Yadav_Haznitrama_Song_Oh_Augenstein_2024}, additional fine-tuning processes \cite{Masoud_Liu_Ferianc_Treleaven_Rodrigues_2024}, enhanced prompting strategies \cite{AlKhamissi_ElNokrashy_Alkhamissi_Diab_2024}, and new benchmarks \cite{Santurkar_Durmus_Ladhak_Lee_Liang_Hashimoto_2023}. These interventions are framed under the broader category of systems that improve users' acceptance of AI output: ``cultural alignment'' \cite{AlKhamissi_ElNokrashy_Alkhamissi_Diab_2024}, ``cultural inclusion'' \cite{Karamolegkou_Rust_Cui_Cao_Søgaard_Hershcovich_2024, Wadern}, or ``cultural competence'' \citep{bhatt-diaz-2024-extrinsic}.

And yet much of the work on these topics does not ask what ``culture'' is, or seek involvement from the many humanities fields that are defined by that question, though there is increased involvement from the social sciences, especially anthropology \cite{Pawar_Park_Jin_Arora_Myung_Yadav_Haznitrama_Song_Oh_Augenstein_2024,hershcovich_challenges_2022,Adilazuarda_Mukherjee_Lavania_Singh_Aji_O’Neill_Modi_Choudhury_2024} and sociolinguistics \cite{Zhou2025CultureIN}. The result is a narrow definition of culture that rests on the terms of European modernity---e.g. a geographic region or a unifying nationality, language, or racial/ethnic/religious identity. This definition is far more rigid than how humanities scholars understand the term. In the humanities, ``culture'' may be used to refer to ``the way of life of a people, group, or humanity in general,'' but it can also be employed to describe ``the works and practices of intellectual and artistic activity'' that emerge from a particular community or group \citep{Williams_1976, georgeyudiceCulture2014}. These complementary yet distinct definitions are important to keep in mind, since it is not only that, for example, a model’s training data might be produced by people from different cultures, in the first sense of the word, but that the training data is itself an expression of culture, as in the second sense of the word, as is the model itself. 

The distinction between people as cultures, and objects or expressions \textit{of} culture, and our awareness of how both definitions are engaged by generative AI models, is crucial for our understanding of their development and their output. Understanding how training data both reflects cultures, and consists of expressions of those cultures, can lead to more intentional data curation practices. In the commercial arena, we have seen how EleutherAI has developed the Pile with heightened attention to scientific cultures, as evidenced by their inclusion of data from PubMed and arXiv, among others \cite{gao_pile_2020}, and how Pleias recently focused its Common Corpus on cultural heritage (e.g. newspapers, monographs) and under-resourced languages \cite{noauthor_releasing_nodate}. This expanded definition of culture can also lead to new ideas for model development, such as in recent work that treats pretraining datasets like collections or archives \cite{desai2024archival}, examining the spread of books, poetry, and other creative content \cite{Chang_Cramer_Soni_Bamman_2023,dsouzaChatbotCanonPoetry2023,walsh_sonnet_2024}, or in work that considers the creative outputs of large models \cite{Lucy_Bamman_2021}, or trains models with capabilities tuned for historical languages \cite{Yamshchikov_Tikhonov_Pantis_Schubert_Jost_2022}.

An expanded definition of culture also allows us to understand cultural objects, including generative AI models, as expressions of larger power structures (or as active challenges to those structures). 
This understanding can allow us to better identify how and why certain perspectives end up captured in training data and others do not, and to see how models 
are meaningful both for the text and other media that they produce, and as expressions of contemporary tech culture in and of themselves. \textbf{Our suggestion is for AI researchers to clarify which definition of culture they are using in their work: ``culture'' as a group of people, or as the objects or expressions of a group of people.} A more precise definition of culture, one that is calibrated to the particular project, can lead to more intentional choices in the curation of training data, the design of alignment strategies, and the identification of appropriate (and inappropriate) use cases, as well as a more clear-eyed understanding of the perspectives that the text or other media generated by any particular pretrained model, or task for which it is employed, might encode.


\subsection{AI can never be ``representative''}

A recognition of the cultural complexity of historical sources, including training data, also offers a fundamentally different way to approach the issue of bias. As is widely recognized, gaps and biases in training data have contributed to an array of harms perpetuated by AI systems \citep{Buolamwini_Gebru_2018, Noble_2018, Eubanks_2018}. In response, researchers and government agencies alike often call for strategies of bias mitigation, ranging from attempts to ``de-bias'' datasets and AI systems \citep{Bolukbasi_Chang_Zou_Saligrama_Kalai_2016}, to improved documentation of both datasets \cite{ Gebru_Morgenstern_Vecchione_Vaughan_Wallach_III_Crawford_2021} and models \cite{Mitchell_Wu_Zaldivar_Barnes_Vasserman_Hutchinson_Spitzer_Raji_Gebru_2019}, to research into model interpretability \cite{Bhatt_Ravikumar_Moura_2019} and explainability \cite{Danilevsky_Qian_Aharonov_Katsis_Kawas_Sen_2020}. Each of these interventions are necessary, but the issue of bias reflects a deeper structural problem, one that cannot be fixed unless the power differentials that cause structural inequalities are challenged at their source. 

Here 
we might draw from theories and practices from feminist, Black feminist, and archival theory, which offer examples of how to engage with historical sources that contain biases---what humanities researchers would call silences \cite{trouillot_silencing_2015} or absences \cite{Hartman_2008}---that can never be ``de-biased'' or modeled away \cite{bode_why_2020}. Reframing bias not as ``bad data'' but as an effect of unequal structural power redirects our attention to the social, historical, and political conditions that gave rise to the biases, and to the dataset's spaces of indeterminacy and its irrecoverably missing parts \cite{Sherman_Morrison_Klein_Rosner_2024}. This raises new research questions: how can we mark missing data, or amplify the significance of sparse data, rather than pass over these gaps \cite{KoeserLeBlanc2024MissingData}? It also raises questions of  responsibility: are we informed enough about, or connected enough to, the people or cultures that the data represents in order to make these decisions? Have we ``read until we understand?'' \cite{Griffin_2021}. 

Humanities researchers devote significant amounts of time and energy to supplementing our own knowledge, expanding our shared archives, and contributing to collective understanding. But we do not deceive ourselves into thinking that there is ever an end-point to this process. We know we will never have all of the ``data,'' so to speak. Instead, we grapple with the lack of complete representativeness. Scholars of the archive of slavery have spent decades developing strategies for making meaning from damaged archival records. Saidiya Hartman models how to amplify the significance of sparse and violent archival records with narrative detail \cite{Hartman_2020}; Marisa Fuentes posits questions about what might have happened on the basis of historical fact \cite{fuentes_2016}; and Jessica Marie Johnson theorizes the ``null value''---the same as in the relational database---as a means of holding space for the people whose stories cannot be recovered \cite{Johnson_2020}. Some of these approaches are already being applied to the output of biased models (e.g. \cite{noauthor_architect_2023}). We are eager to see them applied to the construction and framing of models themselves. To be clear: to address what is missing from the cultural record, we should not replace the real with the fabricated. We must elevate the accountability, specificity, and reflection that has long accompanied archival work; these methods must be made fundamental to work with data as well.

There is an additional lesson 
here: that any act of dataset creation is a political act. Datasets represent languages, peoples, and cultures, which are necessarily embedded in larger power structures. 
Put another way, there is always a perspective that is encoded in the selection of data, the creation of annotations, the citation of research, the setting of parameters, and the construction of prompts. \textbf{Our suggestion is to abandon the chase for a ``representative'' dataset---which, of course, can never be achieved. Instead, we urge AI researchers to acknowledge that there are always perspectives encoded in their datasets and models, and to take time to document and describe them.} This can lead to more precise motivations for undertaking the work, more accurate assessments of its contributions, and more appropriately-scoped applications.


\subsection{Bigger models are not always better models}

As boyd and Crawford assert, ``Bigger data is not always better data'' \cite{boyd_critical_2012}. This applies to models as well. For the first several years of LLM development, the growth in numbers of parameters was rivaled only by the growth in the number of charts documenting the rising number of those parameters. But the axiom that bigger is better has now met its match. Increasingly, studies have found that more data, more parameters, and more compute may not lead to more accurate output \cite{Zhou_Schellaert_Martínez-Plumed_Moros-Daval_Ferri_Hernández-Orallo_2024,mckenzie_inverse_2024}. These findings have raised interesting technical questions, such as how to generate and use synthetic data for training \citep{pleias2025synth}, as well anxieties that we are ``running out of data'' \cite{robisonOpenAICofounderIlya2024}. For humanities researchers, there is an ongoing tension between these large, ``generalist'' models and the small, curated datasets that are the focus of humanist inquiry, leading to technical questions about whether and how such models could support humanities research. 

These findings also serve to prove a more philosophical point: that the goal of a single universal source of ``intelligence'' is itself a fraught endeavor. As feminist philosophers as early as the seventeenth century have shown \citep{Newcastle_2019, Emecheta_1983, Haraway_1988}, there is no such thing as complete knowledge. Centuries of humanities scholarship has confirmed the asymptotic relationship between greater understanding of the human experience and complete knowledge about it. 
This helps diagnose the problem with the concept of ``Artificial General Intelligence'' (or ``AGI'') and points to the clearest alternative: smaller models, with intentionally-curated datasets, that are trained or fine-tuned for precise tasks. 

In the field of computational humanities, a subfield of DH, 
this work has already begun. Giselle Gonzalez Garcia and Christian Weilbach brought their respective backgrounds in history and computer science to customize an LLM for research in Irish migration studies \cite{gisellegonzalezgarciaIfSourcesCould2023}. Historian Miguel Escobar Varela is fine-tuning a SEA-LION model \cite{ng2025sealion} to transcribe and transliterate one hundred years of scans of Malay-language newspapers \cite{varela2025advancing}. And computational literary scholar Ted Underwood and collaborators are currently training an LLM from scratch on only nineteenth-century English-language writing, so as to be able to run counterfactual experiments on the past \cite{UnderwoodNelsonWilkens2025}. 

The promise of these smaller models and modeling tasks also points to the need for a shift in thinking. Rather than being driven by the ``bigger is better'' agenda that characterizes corporate AI research, we should reestablish a research agenda born of humanistic domain expertise: of the painstaking work in archives and through secondary scholarship that is required to construct the most (but again, not wholly) complete knowledge about a particular author, genre, region, time period, or theme. 
\textbf{We suggest that AI researchers invested in augmenting human understanding begin by asking whether a small model might be better for the task, and if so, involve domain experts.} This will bring us closer to a goal that we all share: of more informed, more accurate, and more precise knowledge. 



\subsection{Not all training data is equivalent}

Until very recently, advances in generative AI have relied upon on vast, heterogeneous pretraining datasets. These draw from a wide range of sources: internet content, public-domain and copyrighted books, scientific articles, and now even synthetic data from other AI models, not to mention a variety (but, to be clear, not the entirety) of texts, images, videos, and sounds in different languages, time periods, and genres \cite{Anil_Dai_Firat_Johnson_Lepikhin_Passos_Shakeri_Taropa_Bailey_Chen_etal._2023,brown_language_2020,nostalgebraist_2022}. Despite this range, pretraining data is often treated as a homogeneous, undifferentiated resource---often likened to ``oil'' \citep{stark2019data}---where individual pieces of data are assumed to be interchangeable and valued primarily for their downstream utility. This shallow approach is further compounded by  ``documentation debt'' \cite{Bandy_Vincent_2021}. The actual contents of pretraining data and why specific sources are chosen are rarely disclosed. This is now widely recognized as an ethical problem as well as a technical one. The influential ``Datasheets for Datasets'' \cite{Gebru_Morgenstern_Vecchione_Vaughan_Wallach_III_Crawford_2021} paradigm introduced a structured approach to documenting machine learning datasets, which has been implemented for high-profile data repositories like HuggingFace. 
Some projects like Olmo \cite{olmo2025olmo3} release coarse-grained information about data sources, proportions, and filtering.
We have also increasingly seen empirical tests that assess how different factors such as dataset age, toxicity, and ``quality'' impact model performance \cite{longpre-etal-2024-pretrainers}.

While these are urgent and necessary interventions, approaching pretraining data from a humanistic perspective offers additional value. For example, the development of richer metadata can help to ensure that datasets more accurately represent the contents they seek to include. For a dataset of books, this might include both computationally and manually derived metadata like the book’s title, publication date, genre, provenance (shadow library, etc.), author, and perhaps even demographic information about the author like race, gender, and geography, as in a project aimed at improving the analysis of millions of volumes of digitized text in the HathiTrust Digital Library \cite{underwood_page-level_2014,Underwood_Kimutis_Witte_2020}. Humanities researchers are also keenly aware that the makeup of one's data---extending to dimensions far beyond toxicity and quality---radically impacts downstream tasks \cite{underwood_page-level_2014,Underwood_Kimutis_Witte_2020}, and have developed several detailed strategies for documenting these dimensions and their impact. For example, digital archeologies of datasets place specific data sources in the broader context of the conditions of their creation \cite{lee_compounded_2021,fyfe_archaeology_2016}, and data narratives situate datasets in social contexts and histories beyond their immediate technical use \cite{pouchard_data_2014}. Another recent approach is the data essay  \cite{post45data, nineteenthdata, responsibledatasets, jca, johd} that describes the data’s historical context, curation, and limitations, as well as more abstract considerations of how the data should be considered in light of specific historical moments, social structures, power dynamics, and cultural phenomena.

Because most AI developers have not cared to understand their training data in depth, how models are influenced by finer features of training data remains underexplored. \textbf{We suggest considering pretraining data's conceptual and cultural characteristics in addition to its technical characteristics and ``impact'' on model performance.} This can help model developers avoid a perpetual cycle of after-the-fact fixes---such as guardrails, Reinforcement Learning from Human Feedback (RLHF), post-hoc content moderation, and other patchwork solutions---and lead to innovation.

\subsection{Openness is not an easy fix}

As an increasing number of open and open-source models have been released for public use, many researchers have moved on from the issues surrounding black-boxed, proprietary, and pay-for-access models. But the question of openness---both what it means and what it implies---remains unresolved. In short: there are no easy fixes when working with objects of culture (which include both training data and models). Here, we introduce some key considerations from the humanities aimed at illustrating how questions of openness, ownership, and access rarely have yes-or-no answers and must be consistently reevaluated as contexts and conditions change.   

Consider debates over the inclusion of copyrighted content in training data. On the surface, this would seem to have an easy answer: individuals should control access to their data, including their creative content, and if they do not want it included in training data, it should not be \cite{Chang_Cramer_Soni_Bamman_2023}. To be clear, preserving the ability of writers and artists to create their art and be fairly compensated for it should remain paramount. But the downstream effects on future scholarship are more difficult to parse. If data from certain sources or cultures is not included in a model, future researchers cannot employ the model to learn about those sources or groups \cite {lekhac2021asian}, and there are examples from computational humanities research that point to how this can be done on a case-by-case basis. For example, \citet{Bamman_Samberg_So_Zhou_2024} use fair use exceptions and an exemption to the Digital Millennium Copyright Act (DMCA) to create one of the largest digitized collections of copyrighted films. The HathiTrust Digital Library, which consists of over 18 million digitized books and grew out of the contentious Google Books project, offers derivative data and virtual environments that enable researchers to access and analyze copyright-restricted materials for educational purposes \cite{hathitrust} (though this will be shut down at the end of 2026 \cite{htrcclosing}). Ironically, the Harvard University Library will finally make available its large collection of out-of-copyright Google-digitized books because of the potential to train AI models on open data \citep{cargnelutti2025institutional}.

Access to data and models from communities that cannot, have not, or choose not to provide consent is another area that must be navigated in community and with care. In the cultural heritage space, there are some success stories. Te Hiku Media has successfully developed the Kaituhi Kaitiakitanga License \cite{KaituhiKaitiakitanga}, preserving M\=aori sovereignty over its language model and training data. The Nweulite Obodo suite of licenses provide African language communities with multiple options to determine how their data should (or should not be) shared \cite{DataScienceLawLab_NwuliteObodoOpenDataLicense_2025}. But these same considerations should apply to community data that is technically ``open.'' For example, fanfiction writers participate in intimate, close-knit communities with with an expectation of privacy, despite publishing their work on the open web. Because of the ease of collecting data from the web, users’ stories and interactions can be gathered and shared even without their knowledge or consent \cite{Dym_Fiesler_2020}.
Similarly, the Documenting the Now project has highlighted the harms that can arise when social media data, especially related to movements like the Black Lives Matter, is archived or shared \cite{jules2018documenting}. 

Further questions arise from the use of synthetic data, and from models' ability to generate ``new'' content that seemingly circumvents real-world harms. Here we must consider the impact of generating what Saidiya Hartman would describe as ``scenes of subjection'' \cite{hartmanScenesSubjectionTerror1997}---text and images that, in restaging scenes of historical violence, even through computational means, introduce questions of our own complicity---even with respect to cultures or phenomena that we might simply seek to learn more about. We must contend with the underlying motivation---and very often, the final use case---of much of the research on minoritized groups: surveillance and, at times, intentional harm. Even if these outcomes are anathema to our own ethics, we must remain aware of any and all downstream harms of our work. \textbf{Our suggestion with respect to this weighty issue is, in many ways, a simple one: ask whose knowledge your data or model is attempting to represent, whether you are authorized to make use of it, and whether your work might result in future harm.} If the answer to the last questions are, respectively, no and yes, redirect course. This will result in more open models that should indeed be open, and fewer that extract others' knowledge without consent.

\subsection{Limited access to compute enables corporate capture}

The capabilities of AI systems have grown tremendously, especially over the past five years. However, much of that growth has been fueled not by new insights, but by the concentration of resources \cite{Whittaker_2021}. While the capabilities of these models are qualitatively different from previous ones, the change that enabled them is almost purely quantitative: massive datasets, massive clusters of compute, and massive neural networks that are well-adapted to take advantage of current hardware. Universities and even governments cannot compete with this corporate hyperscaling. 

For most of the history of computing, there has been a one-way march towards greater capability for lower cost and higher efficiency. But the recent wave of AI development has inverted this curve. The cost to achieve minimal gains in performance is now growing exponentially. The environmental and human costs of mining rare-earth (and not-so-rare) minerals required to fabricate GPUs, the water and energy requirements of data centers, and the toll on both the physical and mental health of those conscripted into tasks like RLHF and content moderation are devastating to those who must endure their direct consequences, as they are to all those working towards more just and livable futures. There are also the financial costs that are, ironically, ruinously expensive for the very companies at the forefront of AI development. This has resulted in an ouroboros of overcapitalized investment in underproven infrastructure and revenue models, while the rest of the world asks why AI is needed at all \cite{widderWatchingGenerativeAI2024}.  


While the term ``late capitalism'' has become a convenient catchphrase to describe the range of injustices we face today---from the minor indignities of data-based ad tracking, to the incontrovertible harms of collusion between governments and corporations---we believe that here again, the humanities can serve as a guide. Historians and theorists of capitalism understand the complexity of its motivations and mechanisms, enabling us to push back in strategic ways and envision alternatives. For example, media scholars Ulises Mejias and Nick Couldry’s theory of data colonialism \cite{Couldry_Mejias_2019} explains how corporations have actively worked to separate the data generated by web browsing and social media sites from the people who produce it, enabling it to become a form of ``surplus value'' that can be monetized by corporations without compensation for those who produce it. Coupled with observations by media and information studies scholars such as \citet{ricaurteDataEpistemologiesColoniality2019} and \citet{grayGhostWorkHow2019}, among others, about how the patterns of exploitative digital labor (data cleaning and labeling, content moderation, and the like) parallel historical patterns of colonial labor exploitation, we can more fully understand how we have arrived at this time and place, when decisions of global significance are being made by five US-based tech CEOs. 


Placing today's technocapitalism within a longer history---as histories and theories of labor, colonialism, and racial capitalism enable us to do---also enriches and strengthens our strategies of resistance. For example, we can look to 
nineteenth-century strategies of mutual aid that emerged among Black Americans, enslaved and free, as they supported those seeking freedom and economic justice \cite{alexandersternColoredConventions, p.gabrielleforemanColoredConventionsProject}, and to historical examples of resistance movements aimed at human exploitation and capitalist infrastructure \cite{roberts2015freedom, ortiz2020filosofía, eddins2021rituals, merchant2023blood}. We can also push ourselves to imagine futures less constrained by or altogether outside of the constraints of capitalism. Examples of Indigenous data futures \cite{harjo_spiral_2019,brown_2023} and viral justice \cite{benjamin_viral_2022} provide glimpses of what is possible when we use our imagination to ``push us beyond the constraints of what we think, and are told, is politically possible'' \citep{benjamin_imagination_2024}. \textbf{Here we urge AI researchers, as we do ourselves, to identify their role in this larger mechanism of capitalism, whether inside an institution of higher education or as corporate researchers.} This will allow us to ask how we might direct our research away from a goal of capitalist ``progress'' and toward futures that benefit more than the very rich few. 

\subsection{AI universalism creates narrow human subjects}

Generative AI has further embroiled society in a ``data episteme'' \cite{Koopman2019-ef}, where we are made to feel that we best know ourselves, and the world, through the acquisition of data. 
This is compounded by the fact that culture itself has become ``content.'' Our daily experiences of ingenuity, mundanity, pleasure, and deviance have become abstracted into data that can be extracted and monetized. 
Generative AI models are now aggressively being promoted as ``intelligent'' or ``trustworthy'', absent any acknowledgment of the economic and political ends for which they are being developed. In sum, the distinctly human qualities of people---real, messy, complicated people---
have been reduced to data (if not missing altogether from what is purported to be a complete model of the world. 

A concern with the moral, expressive, and contemplative aspects of living has long been central to humanistic inquiry \cite{mesthene_1969}. But the current obsession with generative AI has shifted our focus to the synthesis and interpretation of data for control, rather than for our contemplation of and responsibility to the world. This shift can be traced to the rise of European modernity and the ideologies that fueled it---to the view that resources, including people, can be classified and quantified under the guise of rationality, progress, and efficiency
To paraphrase Cedric Robinson, extraction requires abstraction \cite{Robinson2021Black}. In a contemporary context, we might consider Amazon’s practice of algorithmically monitoring its workers in pursuit of ``frictionless,'' ``timely'' deliveries, even as they labor under uncaring, unceasing demands for increased productivity under exploitative working conditions \cite{hrw2025gigtrap}.

The work of philosopher Charles Mills \cite{mills2021illumination} shows how modernity's emphasis on rationality and abstraction is and has always been connected to ideas about race. ``Race,'' Mills writes, ``becomes the signifier of full or diminished humanity, a signifier that is enforced by material practices in a modern racialized world'' \cite{mills2021illumination}. This explains how the ``diminished humanity'' of antiblackness becomes foundational to systematized racial subordination---and more concretely, to the system of slavery underpinned European modernity and imperial expansion. As such, modern definitions of humanity, as well as ideas about personhood, citizenship, and economic subjectivity, are inextricably bound to both antiblackness and imperialism. This line of thinking is what has prompted media theorist Ramon Amaro \cite{amaro2023black} to identify a logic of antiblackness in contemporary machine learning algorithms. He intends this as a rebuke of current practices and as a warning for the future: of the dangers of abstracting patterns detached from ``historical category, stereotype, or a moral imaginary'' into a universal idea of the human.


Amaro is describing the concept of universalism, one of European modernity's most prized hallucinations. The drive for AGI is the latest iteration of universalism's false promise, one that masks its desire for abstraction, power, and control, 
and therefore represents a \textit{lack} of contemplation of or responsibility to the world. Abstraction authorizes the erroneous claim that generative AI models can serve as ``universal knowledge'' machines, rather than what they truly are: statistical enactments of the ideology that ``one can improve---control---a deviant subpopulation by enumeration and classification'' \citep{Hacking2013-or, Chan2025PredatoryData}.
 
This work illuminates generative AI’s reordering of society as a network of relations, one driven by the forces of abstraction, acquisition, and informational and material violence. It is one with a long history, but it is only with the knowledge of this history that generative AI stands a chance of being redirected---at least in part---to serve society and culture \cite{nelson2025fallacies}. \textbf{A suggestion for moving forward is to acknowledge this history and admit how its forces persist into the present, not for the purposes of personal reprimand, but rather to ensure that our future work does not unwittingly replicate this history or the forces that sustain it.} Recognizing this history   
can offer a way to guard against overinflated claims of universalism, and point to 
more liberatory applications of AI research. 


\section{Conclusion: Against Humanities Extraction}
\label{conclusion}

Against the dispiriting backdrop of late capitalism and its fueling of the corporate capture of technical research, it has been heartening to see how the substance if not the development of this research has brought attention to the role and importance of humanistic thinking in ways that we, the authors, have not experienced in our near two-decade long involvement at the forefront of digital and computational humanities research 
\citep{mimnoComputationalHistoriographyData2012, klein2013, goldklein2016debates, brock2018, Johnson2018MarkupB, brock2020distributed, storey2020like, walshGoodreadsClassicsComputational2021b, arnold_distant_2023, Tilton2024Computational, bhyravajjulaMuchDependsWhitespace2025a, haveralsEveryonePrefersHuman2025}. 
The advent of generative AI, and AI more generally, has opened up new possibilities for conversation and collaboration across computing, engineering, and the humanities. Humanities scholars are clearly needed to augment a wide range of computational research conversations, policy debates, and public-facing social and cultural AI products, and many humanists are already practiced in these kinds of collaborations. This has been demonstrated by a trend within many subfields of computer science, including NLP, to turn to histories and theories developed within the humanities for additional context, fresh inspiration, and new ideas \cite{piper-etal-2021-narrative, desai2024archival, croskey-2025, Zhou2025CultureIN}. Recognizing these contributions, some Information Schools have begun to hire researchers with humanities PhDs into faculty positions.  

These are all welcome developments, yet we must remain aware of another risk: the stripping of humanities expertise from the disciplines that produce it, and from the disciplinary structures that ensure it can continue. This is akin to the phenomenon of ``elite capture'' as theorized by 
Ol\'uf\d{\'e}mi T\'a\'iw\`o \cite{taiwo_elite_2022}: the simultaneous valorization of certain groups and their contributions (here ``humanities research'') while taking control of the resources required to sustain them---and therefore taking credit for these contributions while depriving the original group of the ability to chart any future course. How to push back against these forms of capture and control, and general conclusions about the value and significance of humanities research for the present moment in AI, form our final set of observations.

First, we offer some practical advice for interdisciplinary collaborations, including those that FAccT seeks to promote. \textbf{Humanities scholars must be brought to the table as equal partners with technical researchers}. This means collaborating from the initial phases of a project and not simply asking for feedback after the fact, as well as crediting them as coauthors. Technical researchers must also recognize existing institutional asymmetries. Put plainly: humanities researchers teach more and are compensated less. They often have less administrative support and fewer funding mechanisms than STEM faculty. For humanities scholars and technical researchers to meet as equal partners when conducting AI research, their time, participation, students, and staff must be funded like their technical counterparts. This funding must be written into sponsored research and corporate grants, since institutional funding for humanities research is several orders of magnitude less than funding for technical researchers \cite{newfield_humanities_2025}. Furthermore, when new initiatives are announced that make claims to engage with history and culture, either within institutions or at the field level, \textbf{they must include humanities researchers in their leadership structures}. If humanities scholars are not included, it becomes incumbent upon the technical researchers already involved to call this out. Otherwise, whatever the increased attention to or value of the work of the humanities, it will simply not be able to continue at the level required to sustain the production of new ideas and support new researchers who can carry the work forward. Regrettably, the institutional signifier of humanities participation has come most often from the single field that has participated most centrally at FAccT: philosophy. We suggest recognizing that ``ethics'' and ``humanities'' are not equivalent, and that fully considering ethical frameworks (such as ``alignment'') means 
respecting a broader array of humanistic approaches. 

Second, a general observation. Many of the advances in AI research of the past few years have come from stochastic optimization, data management, and GPU coding. It is reasonable to expect that experts in those fields should have a place of influence. But the capabilities and the challenges of contemporary AI also look increasingly familiar to the humanities: \textbf{collating archives, developing theories, surfacing complexity, and generating new arguments, all through the lenses of culture, power, interpretation, and preservation}. It is thus equally reasonable to expect that \textbf{experts in these fields should have similar input}. At the very least, if you are an AI (or FAccT) researcher making claims about language, culture, history, or the human record, or working on expressions of culture such as poetry, novels, art, film, or any creative output at the \textit{very least} orient yourself to the current scholarly discourse and include a citation to at least one or two relevant experts who may have already made (or debunked) related claims. Practically speaking, this will mean looking outside of the ACL Anthology or arXiv. By acknowledging the expertise of humanities scholars and by taking active steps to ensure the continuity of humanities research, we see the best hope of employing generative AI to improve our understanding of the human condition and its range of cultures, and enlisting it in support of a future---or futures---in which all of us can thrive. 

With this work, we seek to spark productive conversations across technical and humanities fields about the uses and limits of generative AI. We write in the spirit of true interdisciplinary collaboration and exchange, and we hope this work serves as an invitation to scholars across varied fields to join together on the basis of mutual respect, working towards a shared goal: resisting the extractive and harmful AI systems that have been presented to us and, when appropriate, designing new systems that can enhance our collective knowledge of human experiences and cultures.

\bibliographystyle{ACM-Reference-Format}
\bibliography{provocations}

\appendix

\section{Endmatter}

\subsection{Generative AI Usage Statement}
No generative AI tools were used in the writing of this paper.

\subsection{Author Contributions}
To be added if accepted. 

\subsection{Acknowledgements}
To be added if accepted. 


\subsection{Competing Interests}
To be added if accepted. 

\subsection{Positionality Statement}
To be added if accepted. 


\subsection{Ethical Considerations Statement}
This paper is primarily a literature review and theory contribution, thus we did not engage in any human subjects research, systems development, or deployment. Our work has been guided by considerations for citational justice, and we have specifically sought to cite the work of scholars from marginalized backgrounds, especially BIWOC and queer people of color.

\subsection{Adverse Impact Statement}
With this work, we seek to spark conversations across technical and humanities fields about the uses and limits of AI. We write in the spirit of true interdisciplinary collaboration and exchange, and we hope this work serves as an invitation to scholars across varied fields to join together on the basis of mutual respect, working towards a goal of building AI systems that enhance our collective knowledge of human experiences and cultures when appropriate, while resisting the extractive and outright harmful AI systems that have been provided to us.

\end{document}